\documentclass[prb,aps,twocolumn,amsmath,amssymb,showpacs]{revtex4-2}
\usepackage{graphicx}
\usepackage{psfrag}
\usepackage{color}
\usepackage{subfigure}
\usepackage{amsmath}
\definecolor{dred}{rgb}{0.7,0.0,0.0}
\allowdisplaybreaks 
\usepackage{array}

\newcommand{\BE}{\begin{equation}}
\newcommand{\BEA}{\begin{eqnarray}}
\newcommand{\EE}{\end{equation}}
\newcommand{\EEA}{\end{eqnarray}}

\newcommand{\e}{{\rm e}}

\begin{document}

%
% Title Page
%

\title{$d$-wave Superconductivity in the Hubbard model on the isotropic triangular lattice and a possibility of the chiral $d+id$ 
pairing as a quasi-stable state}
\author{A. Yamada}
\affiliation{Department of Physics, Chiba University, Chiba 263-8522, Japan}

\date{\today}
%\maketitle

\begin{abstract}

We study $d$-wave superconductivity(SC) in the Hubbard model on the 
isotropic triangular lattice described by 
the hopping parameter $t$ and on-site Coulomb repulsion $U$ 
at zero temperature and half-filling 
using the variational cluster approximation. 
We found that the $d_{xy}$ SC is the ground state below the Mott insulator phase $U/t \lesssim 6$, and  
the energy of chiral $d+id$ SC is slightly higher than the $d_{xy}$ SC.  
The energy difference between the normal and $d_{xy}$ states  
is about $0.02t \sim 0.06t$ for $U/t \simeq 5$. 
This result is semi-quantitatively consistent with the SC  
transition temperature $T_K=3.9$ K of $\kappa$-(BEDT-TTF)$_2$Cu$_2$(CN)$_3$, 
where $t$ is estimated to be about $0.06$ eV, 
and the predicted pairing symmetry $d_{xy}$ agrees with the STM observations.  
The energy difference between the $d+id$ and $d_{xy}$ 
is about $0.01t\sim 0.03t$ for $U/t \simeq 5$ so the transition from $d+id$ to $d_{xy}$, 
or some effects of $d+id$ in $d_{xy}$ phase may be observed in experiments 
for $\kappa$-(BEDT-TTF)$_2$Cu$_2$(CN)$_3$.  

\end{abstract}
 
 
\maketitle
 
% Figures
% fig:model
% fig:sc-form-factors
% fig:potential
% fig:energy-difference

%
%Introduction 
%

\section{Introduction}

%{\it Introduction.---} 
Strong electron correlations lead to interesting phenomena such as superconductivity with various pairing symmetries and purely 
paramagnetic insulator (spin liquid) in low dimensional materials. 
The organic charge-transfer salts $\kappa$-(BEDT-TTF)$_2$Cu$_2$(CN)$_3$
%\cite{lefebvre00,shimizu03,kurosaki05,kanoda3,izawa,ichimaru}
%\cite{bedt-ttf,izawa,ichimaru}
\cite{lefebvre00,shimizu03,kanoda3,kurosaki05,izawa,ichimaru,manna}
 is a good example of such materials, whose  
spin liquid state transits to a superconductor 
with $d_{xy}$ pairing symmetry at $T_K=3.9$ K upon applying pressure
\cite{izawa,ichimaru}.

The superconductivity(SC) of this material was studied theoretically using 
the Hubbard model on the isotropic triangular lattice, which is a simple effective Hamiltonian of this material\cite{kino} 
described by the hopping $t$ and the on-site Coulomb repulsion $U$.  
The variational Monte Carlo\cite{watanabe}, exact solution\cite{clay}, 
renormalization group\cite{tsai,honerkamp}, and density renormalization group\cite{shirakawa} were applied to this model, and 
these studies excluded the $d$-wave SC below the Mott transition point at half-filling. 
These results imply that the SC of this material can not be described by the Hubbard model on the triangular lattice.   

Contrary to these analyses, the studies\cite{kyung,senechal-af,laubach} 
using the cluster dynamical mean filed theories(CDMFT) predicted the $d$-wave SC. 
In the study of the cellular dynamical mean-field theory\cite{kyung}, 2$\times$2 cluster is used as the reference cluster and 8 bath sites are attached to that, and 
it is found that the $d_{x^2-y^2}$ SC is realized below the Mott transition point. 
The analysis of the variational cluster approximation(VCA)\cite{senechal-af} adopted 2$\times$2 and 2$\times$4 clusters as the reference clusters and 
reported that the $d_{x^2-y^2}$ SC is the ground state near the Mott insulator phase and the gap symmetry changes to $d_{xy}$ for 
lower values of $U$.  
However the time-reversal symmetry breaking chiral $d+id$ SC, 
which is an important candidate of the pairing symmetry on the triangular lattice, was not considered in these analyses. 
The chiral $d+id$ state was investigated by VCA using 
2$\times$2 cluster and 6-site triangular clusters\cite{laubach} as the reference clusters, and it is found 
that the 2$\times$2 site analysis yields a strong preference for the $d_{x^2-y^2}$ SC, 
while the ground state is the chiral $d+id$ SC below the Mott transition point on the 6-site triangular cluster. 
So the pairing symmetries predicted in the CDMFT do not agree with the experiments\cite{izawa,ichimaru}, 
which suggests together with the results of the anayses\cite{watanabe,clay,tsai,honerkamp,shirakawa}, that 
some physics factors not included in the simple Hubbard model may be necessary to understand the SC of $\kappa$-(BEDT-TTF)$_2$Cu$_2$(CN)$_3$ and related materials. Even within the analyses of the CDMFT, the pairing symmetry of the SC phase is still controversial. 

However, in the CDMFT like VCA, it is guaranteed that the results converge to the exact results as the size of the reference cluster 
increases because the electron correlations within the reference cluster are exactly taken into account.  
Therefore we improve the previous studies of the CDMFT\cite{kyung,senechal-af,laubach} 
by increasing the reference cluster size at least 
until the semi-quantitative convergence is observed between two different sizes of the reference clusters. 
After this minimum check, we compare the obtained results with the experiments\cite{izawa,ichimaru} 
to see if the SC of this material is described by this Hamiltonian.   

In this paper we study the $d$-wave SC in the Hubbard model on the isotropic triangular lattice 
by VCA using the 12-site and 14-site clusters in Fig.~\ref{fig:model} at zero temperature and half-filling. 
We found that the results are semi-quantitatively the same for 12-site and 14-site clusters and the ground state is the $d_{xy}$ SC 
below the Mott insulator phase. 
The energy difference between the normal paramagnetic state(PM) and 
the $d_{xy}$ SC is $0.02t \sim 0.04t$ for $U/t \sim 5$. 
The energy of the chiral $d+id$ state is about $0.01t$ higher than that of the $d_{xy}$-wave ground state for $U/t \sim 5 $. 
The $d$-wave SC is not realized above the Mott transition point $7 \lesssim U/t $. 
%These results are not changed very much by the slight difference of the probes, as will be seen later. 
We have confirmed that our overall conclusion is not changed by slight differences of the probes of the symmetry breaking patterns 
by using the superconducting form factors (e) and (f) in addition to (b)$\sim$(d) in Fig.~\ref{fig:model} into our analysis.  

Comparing our results with the experiments of $\kappa$-(BEDT-TTF)$_2$Cu$_2$(CN)$_3$, 
our prediction of the gap symmetry $d_{xy}$ coincides with the analysis of the thermal conductivity\cite{izawa}
and the STM observations\cite{ichimaru}. 
Adopting the estimate $t \sim 0.06$ eV\cite{salt-parameter1,salt-parameter2,salt-parameter3},  
the energy difference $0.02t \sim 0.04t$ between the PM and $d_{xy}$ SC is semi-quantitatively consistent 
with the transition temperature $T_K=3.9$ K, as will be discussed in detail later. 
Thus we consider that the SC of this material is well-described by this model. 
Because of the semi-quantitative convergence of our results and agreement with the experiments, we expect that 
our overall conclusions in VCA are robust with respect to a further increase in the cluster size. 

Our prediction of the ground state pairing symmetry disagrees with 
the preceding analyses of the CDMFT\cite{kyung,senechal-af,laubach} with smaller reference clusters and we shall discuss in detail 
about the origin of the discrepancies later, considering a relation of the shapes of reference clusters and symmetry of the system 
using the concept of the effective potential. 

%, 
%and that analyses of VCA with the reference clusters of appropriate shapes less than twenty sites will be able to predict the subtle physics 
%semi-quantitatively in general.        

%
% Model 
%
\section{Hubbard model on the isotropic triangular lattice and variational cluster approximation}

\begin{figure}
\includegraphics[width=0.36\textwidth]{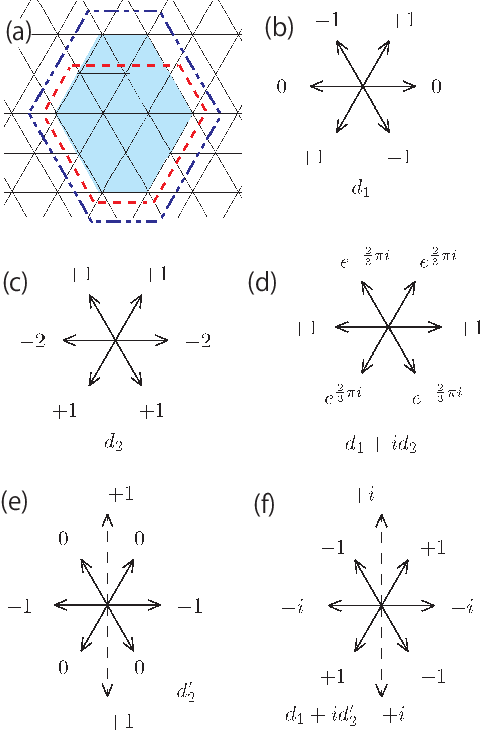}
\caption{(Color online) 
(a) Isotropic triangular lattice. The 12-site cluster in the dotted hexagon and the 14-site shaded cluster in the dash-dotted hexagon are adopted as  
our reference clusters.
(b)$\sim$(f) The real space superconducting form factors used in our analysis. $d_1$ is denoted as $d_{x^2-y^2}$ and 
both $d_2$ and $d'_2$ are denoted as $d_{xy}$ following the terminology on the square lattice. 
%(c) $d_2$ and (e) $d'_2$ 
%have the same symmetric properties under the two space reflections and 
%are topologically equivalent as a probe of symmetry breaking pattern, and 
(e) and (f) are included to confirm that our results are not changed by the subtle difference of the form factors. 
\label{fig:model}\\[-2.0em]}
\end{figure}

%{\it Variational cluster approximation.---}
The Hamiltonian of the Hubbard model on the isotropic triangular lattice reads 
%%%%%%%%%%%%%%%%%%%%%%%%%%%%%%%%%%%%%%%%%%%%%%%%%%%%%%%%%%%%
\begin{align}
H =& -\sum_{i,j,\sigma} t_{ij}c_{i\sigma }^\dag c_{j\sigma}
+ U \sum_{i} n_{i\uparrow} n_{i\downarrow} - \mu \sum_{i,\sigma} n_{i\sigma},
\label{eqn:hm}
\end{align}
%%%%%%%%%%%%%%%%%%%%%%%%%%%%%%%%%%%%%%%%%%%%%%%%%%%%%%%%%%%%
where $t_{ij}=t$ for the solid lines in Fig.~\ref{fig:model}, $U$ is 
the on-site Coulomb repulsion, and $\mu$ is the chemical potential. 
The annihilation (creation) operator for an electron at site $i$ with spin $\sigma$ is denoted as 
$c_{j\sigma}$ ($c_{i\sigma }^\dag$) and $n_{i\sigma}=c_{i\sigma}^\dag c_{i\sigma}$. 
The energy unit is set as $t=1$ hereafter.

% 0 0 246 127

%
% VCA 
%

We use VCA\cite{Senechal00,Potthoff:2003-1,Potthoff:2003} in our analysis. In 
this approach we start with the thermodynamic grand-potential 
$\Omega _{\mathbf{t}}$ written in the form of a functional of the self-energy $\Sigma $ as
\begin{equation}
\Omega _{\mathbf{t}}[\Sigma ]=F[\Sigma ]+\mathrm{Tr}\ln(-(G_0^{-1}-\Sigma )^{-1}).
\label{eqn:omega}
\end{equation}%
In Eq. (\ref{eqn:omega}), $G_0$ is the non-interacting Green's function of $H$, $F[\Sigma ]$ is the Legendre 
transform of the Luttinger-Ward functional\cite{lw}, and the index $\mathbf{t}$ denotes the explicit dependence of 
$\Omega _{\mathbf{t}}$ on all the one-body operators in $H$. 
The variational principle $\delta \Omega _{\mathbf{t}}[\Sigma ]/\delta \Sigma =0$ leads to the Dyson's equation. 
Eq. (\ref{eqn:omega}) gives the exact grand potential of $H$ for the exact self-energy of $H$, which satisfies Dyson's equation. 

All Hamiltonians with the same interaction part share the same functional form of $F[\Sigma ]$, and using that property 
we can evaluate $F[\Sigma ]$ for $\Sigma$ of a simpler Hamiltonian $H'$ by exactly solving it. 
In VCA, we divide the original infinite cluster into the identical clusters, referred to as the reference cluster, 
that tile the infinite lattice, and construct $H'$ from $H$ by removing the hopping interactions between these clusters. 
Then writing the Eq. (\ref{eqn:omega}) for $H'$, and substituting it from $\Omega _{\mathbf{t}}[\Sigma ]$ of $H$, 
we obtain 
\begin{eqnarray}
\Omega _{\mathbf{t}}[\Sigma ]= \Omega' _{\mathbf{t'}}[\Sigma ] &+& \mathrm{Tr}\ln(-(G_0^{-1}-\Sigma )^{-1})
\nonumber \\
&-&\mathrm{Tr}\ln(-(G'_0{}^{-1}-\Sigma )^{-1}),
\label{eqn:omega3}
\end{eqnarray}%
where $G'_0$ is the non-interacting Green's function of $H'$ and $\mathbf{t}'$ denotes all the one-body operators in $H'$. 
In Eq. (\ref{eqn:omega3}) we evaluate $\Omega' _{\mathbf{t'}}[\Sigma ]$ for $\Sigma$ of $H'$ by exactly solving it, then 
$\Omega _{\mathbf{t}}[\Sigma ]$ becomes 
a function of $\mathbf{t}'$ expressed as 
\begin{equation}
\Omega _{\mathbf{t}}(\mathbf{t}')=\Omega' _{\mathbf{t'}} - \int_C{\frac{%
d\omega }{2\pi }} \e^{ \delta \omega} \sum_{\mathbf{K}}\ln \det \left(
1+(G_0^{-1}\kern-0.2em -G_0'{}^{-1})G'\right),
\nonumber
\end{equation}%
where $\Omega' _{\mathbf{t'}}$ is the exact grand potential of $H'$ and 
the functional trace has become an integral over the diagonal variables 
(frequency and super-lattice wave vectors) of the logarithm of the determinant over intra-cluster indices. 
The frequency integral is carried along the imaginary axis and $\delta \rightarrow + 0$. 

The variational principle $\delta \Omega _{\mathbf{t}}[\Sigma ]/\delta \Sigma =0$ is reduced to the stationary condition 
$\delta \Omega _{\mathbf{t}}(\mathbf{t}') /\delta \mathbf{t}' = 0$, and its solution and the 
exact self-energy of $H'$ at the stationary point, denoted as $\Sigma^{*}$, are the approximate grand-potential 
and self-energy of $H$ in VCA. Physical quantities, such as expectation values of one-body operators, 
are evaluated using the Green's function $G_0{}^{-1}-\Sigma^{*}$. 
In VCA, the restriction of the space of the self-energies $\Sigma$ into that of $H'$ 
is the only approximation involved and short-range correlations within the reference cluster are exactly taken into account 
by exactly solving $H'$. 
A possible symmetry breaking is investigated by including in $H'$ the corresponding Weiss field that will be 
determined by minimizing the grand-potential $\Omega_{\mathbf{t}}$. 

In our analysis, the 12-site and 14-site clusters in Fig.~\ref{fig:model} 
are used as the reference clusters to set up the cluster Hamiltonian $H'$. 
We refer to these clusters as 12D and 14D hereafter. 
Within these reference clusters, every site is connected to at least three other sites.

To study the superconductivity, we include the Weiss field Hamiltonian 
\begin{eqnarray}
H_{\rm SC}&=& z \sum_{ij} 
\{ \Delta_{ij} c_{i \downarrow}  c_{j \uparrow} +  \Delta^*_{ij} c_{j \uparrow}^\dag  c_{i \downarrow}^\dag \}
\label{eqn:weiss}
\end{eqnarray}
into $H'$, where $z$ is treated as a variational parameter, and adopt the Nambu formalism  
$c_{j\uparrow} = \tilde{c}_{j\uparrow}$ and 
$c_{i\downarrow } = \tilde{c}_{i\downarrow }^\dag $.

We classify the real-space superconducting form factors $\Delta_{ij}$ between the nearest-neighbor sites into the 
irreducible representations of the invariant group of regular hexagon $C_{6v}$, 
and take as (b) $d_1$ and (c) $d_2$ in Fig.~\ref{fig:model}. 
$C_{6v}$ consists of six rotations (one is identity) and six mirror reflections, and 
%The Weiss fields $d_1$ and $d_2$ belong to the two dimensional irreducible representation $E_2$ of 
%$C_{6v}$\cite{tanabe}, and 
$d_1$ is anti-symmetric under two of the six mirror reflections while $d_2$ is symmetric under these reflections.  
As for the time-reversal symmetry breaking chiral superconducting state, 
we take the combination (d) $i(\sqrt{3}d_1+id_2)/2$ and denote as $d_1+id_2$ hereafter.
   
%In the STM experiments of $\kappa$-(BEDT-TTF)$_2$Cu$_2$(CN)$_3$\cite{ichimaru}, 
In the STM experiments\cite{ichimaru}  
the two line nodes of the gap are ${\pi}/{4}$ from the axes of the two reciprocal lattice vectors, which 
indicates $d_2$, whose two line nodes lie between the basic lattice vectors in the real space. 
Also, the observed gap\cite{ichimaru} is symmetric under two mirror reflections.   
In the case of $d_1$, one of the two line nodes lies along the basic lattice vectors, while the other is perpendicular to 
that, which does not agree with the result of STM. 

To confirm that subtle differences of the choice of the Weiss field do not change our results, we also consider 
(e) $d'_{2}$ in Fig.~\ref{fig:model}, which involves next-nearest neighbor interactions, and (f) $d_1 + i d'_{2}$. 
$d'_2$ has the same symmetric property as $d_2$ and also probes the SC in the $d_2$ direction. 
The two chiral states $d_1+id_2$ and $d_1 + i d'_{2}$ are topologically equivalent since both the real and imaginary 
parts change the sign twice in a similar manner by $2\pi$ rotation.
Because the terminology of the gap $d_1 \equiv d_{x^2-y^2}$ and $d'_2 \equiv d_{xy}$ on the square lattice 
are widely used in the previous studies\cite{izawa,ichimaru,kyung,senechal-af,laubach},  
we also denote $d_1$ as $d_{x^2-y^2}$, and both $d_2$ and $d'_2$ as $d_{xy}$.

With this set up, $\mathbf{t}'$ is reduced to the Weiss parameter $z$ in Eq. (\ref{eqn:weiss}) and the cluster chemical potential $\mu'$, 
where $\mu'$ should be included for the thermodynamic consistency\cite{aichhorn},  
and the stationary condition 
$\delta \Omega _{\mathbf{t}}(\mathbf{t}') /\delta \mathbf{t}' = 0$, 
is solved by searching 
the stationary point of $\Omega(\mu', z)$, which we denote as the grand-potential per site. 
During the search, the chemical potential of the system $\mu$ is also adjusted so that the electron 
density $n$ is equal to 1 within the accuracy of $10^{-5}$. 
The energy per site is given by $E=\Omega+\mu n$ where $\Omega$ is the value of $\Omega(\mu', z)$ at the 
stationary point.  
In general, a stationary solution with $z \neq 0$ corresponding to the superconducting state and 
that with $z = 0$ corresponding to the normal paramagnetic state 
are obtained, and these energies are compared to determine the ground state.

\section{RESULTS}

%{\it The ground state.---}
Before the analysis of the SC, we compute the Mott transition point for the PM. 
To examine the gap we calculated the density of state per site 
\begin{eqnarray}
%D(\omega)= \lim_{\eta \rightarrow 0}  \int%_{BZ}
%{\frac{%
%d^2 k }{(2\pi)^2 }}\frac{1}{n_c}\sum_{\sigma, a=1}^{n_c}\{ -\frac{1}{\pi} \mathrm{Im}G_{a\sigma}(k, \omega+i\eta) \}
D(\omega)= \lim_{\eta \rightarrow 0}  \int%_{BZ}
{\frac{%
d^2 k }{(2\pi)^2 }} \{ -\frac{1}{\pi} \mathrm{Im}G_{a\sigma}(k, \omega+i\eta) \}
\label{eqn:dos}
\end{eqnarray}
imposing $z=0$, where  
%In Eq. (\ref{eqn:dos}), 
the $k$ integration is over the corresponding Brillouin zone, and $\eta \rightarrow 0$ limit is 
evaluated using the standard extrapolation technique, 
and obtained that the Mott transition point $U_M$ is $U_M=6.3$ on 
12D\cite{laubach,atsushi-triangle1,atsushi-triangle2}, and $U_M=5.4$ on 14D. 

Next we consider the SC for $U_M \lesssim U$. 
At $U=9$ and $U=12$, we found that $\Omega(\mu', z)$ is monotonically increasing as a function of $z$ near the half-filling 
and no SC solutions are obtained. 
At $U=7$, we found the stationary solution for all $d_1$, $d_2$, $d_1+id_2$, $d_1 + i d'_{2}$, and $d'_2$ on both 12D and 14D, 
and their energies are degenerate with the PM within about $10^{-4}t \sim 10^{-5}t$. 
Therefore we consider that the SC is not realized for $U_M \lesssim U$. 

\begin{figure}
\includegraphics[width=0.42\textwidth]{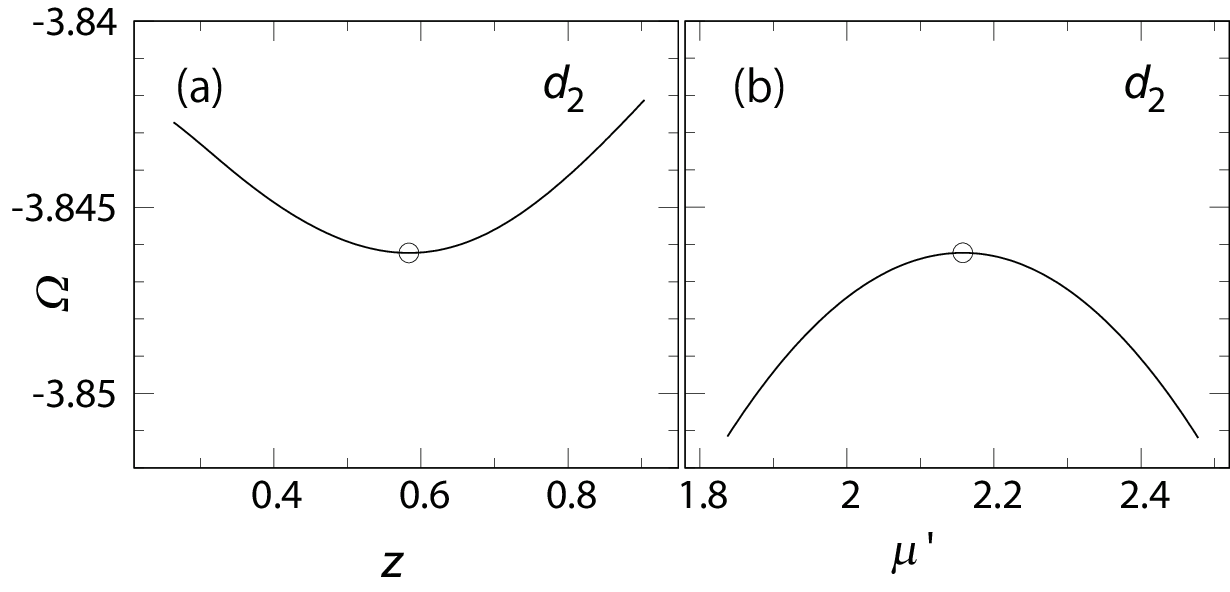}
\caption{(Color online) 
The grand potential per site $\Omega(\mu',z)$ as a function of
 (a) z and (b) $\mu'$ computed for $d_2$ at $U=4$ on 14D by VCA. 
We set $\mu=2.7223019$. 
The circle corresponds to the stationary solution at half-filling. 
In (a) $\mu'$ is set to be the stationary solution value, and 
in (b) $z$ is set to be the stationary solution value.  
\label{fig:14d2}\\[-1.5em]}
\end{figure}

\begin{figure}
\includegraphics[width=0.42\textwidth]{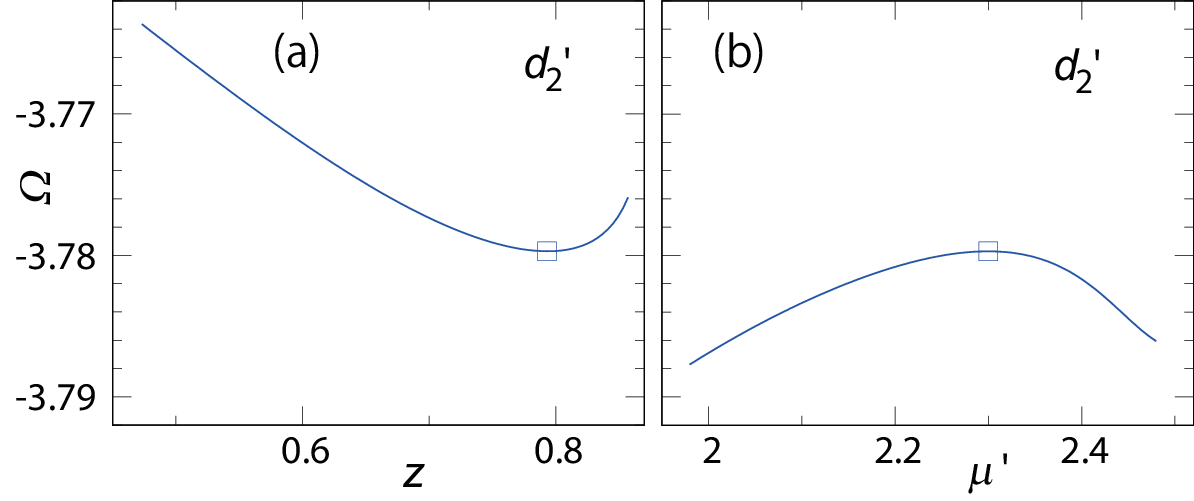}
\caption{(Color online) 
The grand potential per site $\Omega(\mu',z)$ as a function of
 (a) z and (b) $\mu'$ computed for ${d'}_2$ at $U=4$ on 14D by VCA. 
We set $\mu=2.6329068$. 
The circle corresponds to the stationary solution at half-filling. 
In (a) $\mu'$ is set to be the stationary solution value, and 
in (b) $z$ is set to be the stationary solution value.  
\label{fig:14dxy}\\[-1.5em]}
\end{figure}

\begin{figure}
\includegraphics[width=0.42\textwidth]{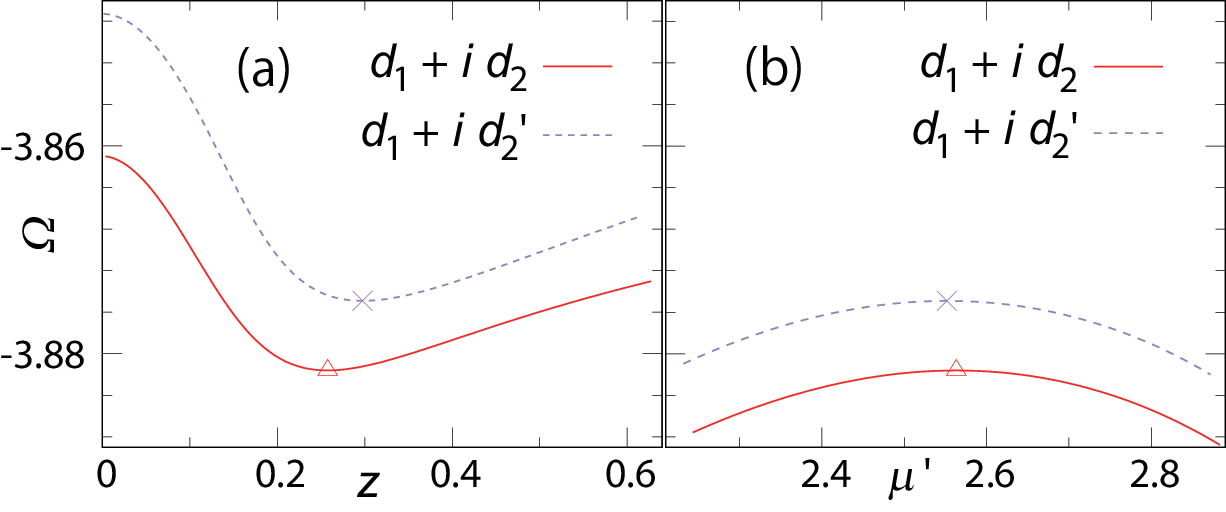}
\caption{(Color online) 
The grand potential per site $\Omega(\mu',z)$ as a function of
 (a) z and (b) $\mu'$ computed for $d_1+id_2$ and $d_1+id'_2$ at $U=4$ on 14D by VCA. 
We set $\mu=2.773955$ for $d_1+id_2$, and $\mu=2.7605867$ for $d_1+id'_2$. 
The triangles and crosses indicate the stationary points satisfying $n=1$.  
In (a) $\mu'$ is set to be the stationary solution value, and 
in (b) $z$ is set to be the stationary solution value.  
\label{fig:14idx2y2-d2-mod}\\[-1.5em]}
\end{figure}

Next we consider the SC for $U \lesssim U_M$. 
Figs.~\ref{fig:14d2}--\ref{fig:14idx2y2-d2-mod} show $\Omega(\mu', z)$ as functions of $z$ and $\mu'$ 
computed by VCA on 14D at $U=4$ for $d_2$, ${d'}_2$, $d_1+i{d}_2$ and $d_1+i{d'}_2$. 
In these figures, the marks show the stationary point satisfying the half-filling condition $n=1$. 
In figures (a), we fix $\mu'$ to be the value of the stationary solution, and in (b) we fix $z$ to be the value of the stationary solution. 
The values of $\mu$ are given in the caption. 
Similarly, we found the stationary solutions satisfying $n=1$ for $d_2$, $d_1+id_2$, and $d_1 + i d'_{2}$ 
at $U=1,2,3,4,5$,and $6$ on 12D and 14D. 
As for $d'_2$, we found the stationary solutions 
for $1 \leq U \leq 6$ on 14D and $4\leq U \leq 6$ on 12D, but 
we were not able to obtain the stationary solutions 
for $U \leq 3$ on 12D because $\Omega(\mu', z)$ becomes discontinuous. 
In general, long range correlations become more important for smaller $U$, 
and 12D cluster is slightly small to simulate $d'_2$ for $U \leq 3$.  
On 12D we continued the stationary point search up to larger values $z\simeq 1.6$. Sometimes these searches are terminated by the discontinuity of the grand potential 
due to the change of the electron numbers of the cluster ground state. 
By these searches we confirmed that the obtained solutions are unique.

%\begin{figure}
%\includegraphics[width=0.47\textwidth]{pot-z-v3}
%\caption{(Color online) 
%The grand potential per site $\Omega(\mu',z)$ as a function of
%z computed for $d_1$ at $U=4$ on 14D by VCA. 
%We set $\mu=2.6944759.$  
%The diamond at $z=0$ indicates the stationary points satisfying $n=1$.  
%\label{fig:14d1}\\[1.0em]}
%\end{figure}

%Next we discuss about $d_1$. Fig.~\ref{fig:14d1} shows the grand potential per site $\Omega(\mu', z)$ as functions of $z$ 
%computed by VCA on 14D at $U=4$ for $d_1$. In this figure, the diamond at $z=0$ 
%corresponds to the paramagnetic stationary solution at half-filling.
%Similarly, 

As for $d_1$, we found that $\Omega(\mu',z)$ 
remains monotonically decreasing as we increase $z$, 
until $\Omega(\mu',z)$ becomes discontinuous 
both on 12D and 14D for $1 \leq U \leq 6$. 
Therefore the stationary solution was not obtained for $d_1$.   
A monotonically decreasing behavior of the grand potential is also observed e.g., 
for $s$-wave and $d_{xy}$ pairings for the superconductivity 
in the Hubbard model on the square lattice in VCA 
using $2 \times 2$ cluster\cite{Senechal-lecture-note}.  
Since we will be able to exclude the quasi-stable $s$-wave SC in the 
Hubbard model based on the physics ground, the 
monotonically decreasing behavior does not necessarily imply 
that the reference cluster is too small to simulate 
the corresponding ordered state. 
In our case, we consider that $d_1$ is simulated well on 12D and 14D 
because the other four pairings 
%$d_2$, $d'_2$, $d_1+i d_2$, and $d_1+i{d'}_2$ 
are simulated 
well on these clusters, and the absence of the stationary solution means $d_1$ is not realized for $1 \leq U \leq 6$. 
Rigorously speaking, 
%regardless of the size of the reference cluster, 
whenever stationary solutions are not obtained, 
there always remains a possibility that the reference cluster is not suited to simulate the corresponding states.

\begin{figure}
\includegraphics[width=0.47\textwidth]{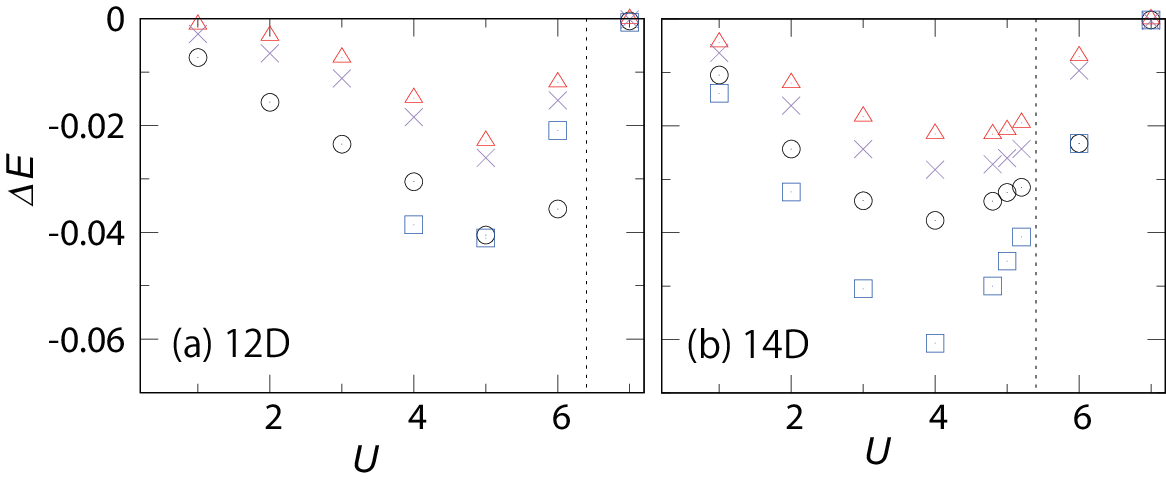}\\%[-1.0em]
\caption{
(Color online)
The energy difference between superconducting and paramagnetic states for 
$d_2$ (circles), ${d'}_2$ (squares), $d_1 + i d_2$(triangles), 
and $d_1 +i d'_2$  (crosses) as functions of $U$ 
computed by VCA on (a) 12D and (b) 14D clusters. The dotted vertical lines correspond to the 
Mott transition points $U_M$. 
As for $d'_2$, the grand potential $\Omega(\mu',z)$ became disconnected for $U=1,2,3$ on 12D and 
stationary solutions were not obtained. 
\label{fig:energy-difference}\\[-2.2em]
}
\end{figure}

Next we analyze the ground state energies of the SC. Fig.~\ref{fig:energy-difference} shows the difference of the energy per site 
between the SC and PM, $\Delta E =  E_{SC}-E_{PM}$ as a function of $U$ computed in VCA on (a) 12D and (b) 14D. 
The dotted lines indicate the Mott transition point $U_M$. 
At $U=5$ on 14D, $\Omega(\mu',z)$ becomes almost flat in the $z$ direction near the stationary point for $d_2$ 
and the determination of the minimum might contain numerical errors, so 
we analyzed the energy differences also for $U=4.8$ and $U=5.2$.   
Taking into account the fact that $d_2$ and $d'_2$ share the same symmetry property and are topologically equivalent, 
the general features are the same for 12D and 14D, 
though the cluster size dependence is not negligible for the magnitudes of the energy differences. 
For $1 \leq U \leq 6 $, $d_{xy}$ ($d_2$ and $d'_2$) SC is the ground state and 
the chiral states $d_1+id_2$ and $d_1 + i d'_{2}$ are energetically disfavored compared to $d_{xy}$. 
The energy difference between the PM and $d_{xy}$ SC is about $0.02t\sim0.06t$ for 
$U \simeq U_M$, and decrease as $U$ decreases. The energy difference between the $d+id$ and $d_{xy}$ 
is about $0.01t\sim 0.03t$ for $U/t \simeq 5$

We compare our results with experiments of $\kappa$-(BEDT-TTF)$_2$Cu$_2$(CN)$_3$, which exhibits a transition 
from spin liquid to the $d_{xy}$ SC at $T_K=3.9$ K 
upon applying pressure\cite{lefebvre00,shimizu03,kanoda3,kurosaki05,izawa,ichimaru,manna}. 
Our prediction of the pairing symmetry is $d_{xy}$ and agrees with the experiments.  
For more quantitative comparisons, we assume that applying the pressure has the effect of 
increasing $t$ without affecting $U$, thus decreasing $U/t$. 
Then the Mott insulator region $5 \lesssim U/t \lesssim 7 \sim 8$ slightly above the SC 
is the candidate for the spin liquid, 
so we adopt the estimate $5 \lesssim U/t \lesssim 8$\cite{salt-parameter1,salt-parameter2,salt-parameter3}.  
Our analysis excludes the estimate $12 \lesssim U/t \lesssim 15$\cite{salt-parameter3} since the pressure in the experiments 
will not be able to push this region of $U/t$ down to the $d_{xy}$-wave SC phase $U/t \lesssim 7$.

The energy difference between the PM and $d_{xy}$ SC is about $0.02t \sim 0.06$ for $U \simeq U_M$. 
Even though the precise determination of the transition temperature requires the analysis of the entropy factor and finite temperature free energy,  
the transition temperature will have to be sizably lower than this energy difference to realize this $d_{xy}$ SC, since the 
entropy factor will be larger for the higher energy PM 
compared to the SC, and tend to increase the probability of the appearance of the PM.  
Assuming $t \simeq 0.06$ eV\cite{salt-parameter1}, 
this energy difference corresponds to $14 \sim 42$~K, and the transition temperature $T_K=3.9$ K in the experiment is sizably lower than this value. 

There exists fully-gapped $d+id$ slightly higher than the nodal gap $d_{xy}$, and might be observed as a quasi-stable state. 
It is an interesting challenge to observe the transition from $d+id$ to $d_{xy}$, 
or some effects of $d+id$ in $d_{xy}$ phase in experiments of this material. 

Finally we discuss about the origin of the discrepancies between the preceding analyses\cite{kyung,senechal-af,laubach} and ours.   
%The importance of the cluster shape and symmetry of the system in CDMFT was pointed out in Ref. \onlinecite{laubach}, 
%especially for smaller reference clusters, since the electron correlations within the reference clusters are 
%exactly taken into account in the CDMFT.  
%It is pointed out in Ref. \onlinecite{laubach} that 
In general it is important in the CDMFT with relatively small reference clusters to choose appropriate shapes of reference clusters taking into account the symmetry of the system because the electron correlations within the reference clusters are exactly taken into account.   
In the case of isotropic triangular lattice, there is $C_{6v}$ symmetry, and 
the order parameters of $d_{x^2-y^2}$ and $d_{xy}$ transform according to the two dimensional irreducible representation $E_2$ of $C_{6v}$\cite{tanabe}. 
Because of this property, they mix in the singlet effective potential classified by the group symmetry, which leads to the possibility of the formation of $d+id$. 
Also, the two states $d_{x^2-y^2}$ and $d_{xy}$ are degenerate up to the fourth-order expansion of the effective potential 
in terms of the order parameters, and sixth-order terms are necessary to solve this degeneracy\cite{sigrist-ueda}.  
The electron correlations within the square shape 2$\times$2 and 2$\times$4 clusters used in the preceding studies\cite{kyung,senechal-af,laubach} 
would not be able to create well the effective potential of this symmetry property and these analyses could not 
predict accurately the stable SC pairing state on the isotropic triangular lattice. 
The 6-site triangular cluster used in Ref. \onlinecite{laubach} and our 12D 
do not keep $C_{6v}$ symmetry, but they keep $C_{3v}$ symmetry, and the situations are the same to the case of $C_{6v}$ in the sense that 
the order parameters of $d_{x^2-y^2}$ and $d_{xy}$ transform according to the two dimensional irreducible representation 
$E$ of $C_{3v}$\cite{tanabe}, and they mix in the singlet effective potential, leading to the possibility of the formation of $d+id$.

The origin of the discrepancies between the results of the 6-site triangular cluster in Ref. \onlinecite{laubach} and ours will be the difference of the cluster size.  
Our results suggest that, even though electron correlations within 4-site or 8-site clusters may be adequate to predict some physics quantities in VCA,   
clusters of more than ten sites are necessary to predict delicate physics property 
like pairing symmetry of the SC on the isotropic triangular lattice. 
As for the cluster size dependence, we have checked that our results of 12D are semi-quantitatively the same to those of 14D. 
Our 14D does not keep $C_{3v}$ symmetry, but the results of VCA converge to the exact results as the size of the reference cluster increases 
regardless of the detailed shapes of the cluster, and in fact they did for 12D and 14D.   
Since the physics discussed here seems to be delicate, we have also checked that our overall conclusions are not changed by 
the slight difference of the choice of the Weiss fields. 
Because of these checks and the semi-quantitative agreement between the experiments and our results, 
we expect that our results are semi-quantitatively robust with respect to a further increase in the reference cluster size.

\section{SUMMARY} 

%{\it Summary and perspectives.---}

We have studied the $d$-wave superconductivity in the Hubbard model on the isotropic triangular lattice by VCA 
at $n=1$. We have improved the preceding analyses by increasing the reference cluster size while paying attention to the 
cluster shape and symmetry of the system. We found that the results are the semi-quantitatively 
the same for the 12-site and 14-site clusters and the ground state is the $d_{xy}$ SC below the Mott transition point 
$U/t \lesssim 7$. The $d$-wave SC is not realized for $7 \lesssim U$. 
Our prediction of the pairing symmetry $d_{xy}$ agrees 
with the electronic thermal conductivity\cite{izawa} and STM experiments\cite{ichimaru} of $\kappa$-(BEDT-TTF)$_2$Cu$_2$(CN)$_3$, and 
adopting the estimate $5 \lesssim U/t \lesssim 8$ and $t=0.06$ eV\cite{salt-parameter1}, 
our result is semi-quantitatively consistent with the SC transition temperature $T_K=3.9$ K of this material.  
We also found that there exists the chiral $d+id$ SC slightly above the ground state $d_{xy}$ SC, 
and it is an interesting challenge to observe the transition from $d+id$ to $d_{xy}$, 
or some effects of $d+id$ in $d_{xy}$ phase in the experiments of $\kappa$-(BEDT-TTF)$_2\mathrm{X}$.

\section*{ACKNOWLEDGMENT}

I thank H.~Fukazawa, J.~Goryo, H.~Kurasawa, H.~Nakada, T.~Ohama, and Y.~Ohta for useful discussions. 
Parts of numerical calculations were done using the computer facilities of 
the IMIT at Chiba University, ISSP, and Yukawa Institute.


\begin{thebibliography}{99}

%--- kappa-(BEDT-TTF)2Cu[N(CN)2]Cl Mott transition ------------------%
%\bibitem{bedt-ttf}
%S. Lefebvre, P. Wzietek, S. Brown, C. Bourbonnais, D. J{\'e}rome, C. M{\'e}zi{\`e}re, M. Fourmigu{\'e}, and P. Batail, 
%Phys. Rev. Lett. {\bf 85}, 5420 (2000);
%Y. Shimizu, K. Miyagawa, K. Kanoda, M. Maesato, and G. Saito, 
%Phys. Rev. Lett. 
%({\it ibid.}) 
%{\bf 91}, 107001 (2003);
%F. Kagawa, T. Itou, K. Miyagawa, and K. Kanoda, 
%Phys. Rev. B {\bf 69}, 064511 (2004);
%Y. Kurosaki, Y. Shimizu, K. Miyagawa, K. Kanoda, and G. Saito, 
%Phys. Rev. Lett. 
%({\it ibid.}) 
%{\bf 95}, 177001 (2005).

\bibitem{lefebvre00}
S. Lefebvre, P. Wzietek, S. Brown, C. Bourbonnais, D. J{\'e}rome, C. M{\'e}zi{\`e}re, M. Fourmigu{\'e}, and P. Batail, 
Mott Transition, Antiferromagnetism, and Unconventional Superconductivity in Layered Organic Superconductors, 
Phys. Rev. Lett. {\bf 85}, 5420 (2000).

\bibitem{shimizu03}
Y. Shimizu, K. Miyagawa, K. Kanoda, M. Maesato, and G. Saito, 
{\it et al.}, 
Spin Liquid State in an Organic Mott Insulator with a Triangular Lattice, 
Phys. Rev. Lett. {\bf 91}, 107001 (2003).


\bibitem{kanoda3}
F. Kagawa, T. Itou, K. Miyagawa, and K. Kanoda, 
Transport criticality of the first-order Mott transition in the quasi-two-dimensional organic conductor $\kappa$-(BEDT-TTF)$_2$Cu[N(CN)$_2$]Cl, 
Phys. Rev. B {\bf 69}, 064511 (2004). 

\bibitem{kurosaki05}
Y. Kurosaki, Y. Shimizu, K. Miyagawa, K. Kanoda, and G. Saito, 
Mott Transition from a Spin Liquid to a Fermi Liquid in the Spin-Frustrated Organic Conductor $\kappa$-ET$_2$Cu$_2$(CN)$_3$
Phys. Rev. Lett. {\bf 95}, 177001 (2005).

\bibitem{izawa}K. Izawa, H. Yamaguchi, T. Sasaki and Y. Matsuda, 
Superconducting Gap Structure of $\kappa$-(BEDT-TTF)$_2$Cu(NCS)$_2$ Probed by Thermal Conductivity Tensor, 
Phys. Rev. Lett. {\bf 88} 027002 (2001).

\bibitem{ichimaru} 
K. Ichimura and K. Nomura, 
d-wave Pair Symmetry in the Superconductivity of $\kappa$-(BEDT-TTF)$_2$X, 
J. Phys. Soc. Jpn. {\bf 75}, 051012 (2006).  

\bibitem{manna} R.S. Manna, M. de Souza, A. Br\"uhl, J.A. Schlueter, and M. Lang,
Lattice Effects and Entropy Release at the Low-Temperature Phase Transition in the Spin-Liquid Candidate $\kappa$-(BEDT-TTF)$_2$Cu$_2$(CN)$_3$, 
{\prl} {\bf 104}, 016403 (2010).

\bibitem{kino}
H. Kino and H. Fukuyama, 
Phase Diagram of Two-Dimensional Organic Conductors: (BEDT-TTF)$_2$X, 
J. Phys. Soc. Jpn. {\bf 65}, 2158 (1996).

%spin liquid

%\bibitem{imada} H. Morita, S. Watanabe, and M. Imada, 
%Nonmagnetic insulating states near the Mott transitions on lattices with geometrical frustration and implications for $\kappa$-(BEDT-TTF)$_2$Cu$_2$(CN)$_3$, 
%J. Phys. Soc. Jpn. {\bf 71}, 2109 (2002).

\bibitem{watanabe} T. Watanabe, H. Yokoyama, Y. Tanaka, and J.-i. Inoue, 
Superconductivity and a Mott transition in a Hubbard model on an anisotropic triangular lattice,
J. Phys. Soc. Jpn. {\bf 75}, 074707 (2006).

\bibitem{clay} R.T. Clay, H. Li, and S. Mazumdar, 
Absence of Superconductivity in the Half-Filled Band Hubbard Model on the Anisotropic Triangular Lattice, 
{\prl} {\bf 101}, 166403 (2008).



%SC RG
%\bibitem{rg} 
%S.W. Tsai and J.B. Marston, 
%Weak-coupling functional renormalization-group analysis of the Hubbard model on the anisotropic triangular lattice, 
%Can. J. Phys. 79, 1463 (2001); 
%C. Honerkamp, 
%Phys. Rev. B {\bf 68}, 104510 (2003). 

\bibitem{tsai} 
S.W. Tsai and J.B. Marston, 
Weak-coupling functional renormalization-group analysis of the Hubbard model on the anisotropic triangular lattice, 
Can. J. Phys. 79, 1463 (2001). 

\bibitem{honerkamp} 
C. Honerkamp, 
Instabilities of interacting electrons on the triangular lattice, 
Phys. Rev. B {\bf 68}, 104510 (2003). 

\bibitem{shirakawa}
T. Shirakawa, T. Tohyama, J. Kokalj, S. Sota, and S. Yunoki, 
Ground-state phase diagram of the triangular lattice Hubbard model by the density-matrix renormalization group method, 
Phys. Rev. B 96, 205130 (2017).

\bibitem{kyung} B. Kyung and A.-M. S. Tremblay, 
Mott Transition, Antiferromagnetism, and $d$-Wave Superconductivity in Two-Dimensional Organic Conductors,
{\prl} {\bf 97}, 046402 (2006).

\bibitem{senechal-af} P. Sahebsara and D. S\'en\'echal, 
Antiferromagnetism and Superconductivity in Layered Organic Conductors: Variational Cluster Approach,
{\prl} {\bf 97}, 257004 (2006).

\bibitem{laubach} M. Laubach, R. Thomale, C. Platt,W. Hanke, and G. Li, 
Phase diagram of the Hubbard model on the anisotropic triangular lattice, 
Phys. Rev. B {\bf 91}, 245125 (2015).


%\bibitem{salt-parameter1}
%T. Komatsu, N. Matsukawa, T. Inoue, and G. Saito, 
%J. Phys. Soc. Jpn. {\bf 65}, 1340 (1996);
%H.C. Kandpal, I. Opahle, Y.-Z. Zhang, H.O. Jeschke, and R. Valenti, 
%{\prl} {\bf 103}, 067004 (2009);
%K. Nakamura, Y. Yoshimoto, T. Kosugi, R. Arita, and M. Imada, 
%J. Phys. Soc. Jpn. {\bf 75}, 074707 (2009); 
%K. Nakamura, Y. Yoshimoto, and M. Imada, 
%Phys. Rev. B {\bf 86}, 205117 (2012). 

\bibitem{salt-parameter1}
T. Komatsu, N. Matsukawa, T. Inoue, and G. Saito, 
Realization of Superconductivity at Ambient Pressure by Band-Filling Control in $\kappa$-(BEDT-TTF)$_2$Cu$_2$(CN)$_3$, 
J. Phys. Soc. Jpn. {\bf 65}, 1340 (1996). 
 
\bibitem{salt-parameter2}
H.C. Kandpal, I. Opahle, Y.-Z. Zhang, H.O. Jeschke, and R. Valenti, 
Revision of Model Parameters for $\kappa$-Type Charge Transfer Salts: An $Ab Initio$ Study, 
{\prl} {\bf 103}, 067004 (2009). 

\bibitem{salt-parameter3}
K. Nakamura, Y. Yoshimoto, T. Kosugi, R. Arita, and M. Imada, 
$Ab initio$ derivation of low-energy model for $\kappa$-ET type organic conductors
J. Phys. Soc. Jpn. {\bf 75}, 074707 (2009); 
K. Nakamura, Y. Yoshimoto, and M. Imada, 
$Ab initio$ two-dimensional multiband low-energy models of EtMe$_{3}$Sb[Pd(dmit)$_{2}$]$_{2}$ and $\kappa$-(BEDT-TTF)$_2$Cu(NCS)$_2$ with comparisons to single-band models 
Phys. Rev. B {\bf 86}, 205117 (2012).  

%VCA
\bibitem{Senechal00}
D.~S\'en\'echal, D. Perez, and M. Pioro-Ladri\'ere, 
Spectral Weight of the Hubbard Model through Cluster Perturbation Theory, 
Phys. Rev. Lett. {\bf 84}, 522 (2000); 
D.~S\'en\'echal, D. Perez, and D. Plouffe, 
%Cluster perturbation theory for Hubbard models, 
Phys. Rev. B {\bf 66}, 075129 (2002).

\bibitem{Potthoff:2003-1}
M.~Potthoff, M.~Aichhorn, and C.~Dahnken, 
Variational Cluster Approach to Correlated Electron Systems in Low Dimensions, 
Phys. Rev. Lett. {\bf 91} 206402 (2003); 
C.~Dahnken, M.~Aichhorn, W. Hanke, E. Arrigoni, and M.~Potthoff, 
Variational cluster approach to spontaneous symmetry breaking: The itinerant antiferromagnet in two dimensions, 
Phys. Rev. B {\bf 70}, 245110 (2004).

\bibitem{Potthoff:2003} 
M. Potthoff, 
Self-energy-functional approach to systems of correlated electrons, 
Eur. Phys. J. B \textbf{32}, 429 (2003).


\bibitem{lw}   %   mu' 
L. M. Luttinger and J. C. Ward, 
Ground-State Energy of a Many-Fermion System. II, 
Phys. Rev. {\bf 118}, 1417 (1960). 



\bibitem{aichhorn}   %   mu' 
M. Aichhorn, E. Arrigoni, M. Potthoff, and W. Hanke, 
Antiferromagnetic to superconducting phase transition in the hole- and electron-doped Hubbard model at zero temperature, 
Phys. Rev. B {\bf 74}, 024508 (2006). 

%\bibitem{atsushi-triangle1} A. Yamada, 
%Magnetic properties and Mott transition in the Hubbard model on the anisotropic triangular lattice, 
%Phys. Rev. B {\bf 89}, 195108 (2014); 
%Phys. Rev. B {\bf 90}, 235138 (2014).

\bibitem{atsushi-triangle1} A. Yamada, 
Magnetic properties and Mott transition in the Hubbard model on the anisotropic triangular lattice, 
Phys. Rev. B {\bf 89}, 195108 (2014).

\bibitem{atsushi-triangle2} A. Yamada, 
Magnetic properties and Mott transition of the Hubbard model for weakly coupled chains on the anisotropic triangular lattice, 
Phys. Rev. {\bf B} 90, 235138 (2014).


\bibitem{Senechal-lecture-note}
D.~S\'en\'echal, arXiv:0806.2690v2: 
Lectures given at the CIFAR-PITP International School on Numerical Methods for Correlated Systems in condensed Matter. 



\bibitem{tanabe} T. Inui, Y. Tanabe, and , Y. Onodera, 
Group theory and its applications in physics, Springer Series in Solid-State Science, 78, (1990).  



\bibitem{sigrist-ueda} 
M. Sigrist and K. Ueda, 
Phenomenological theory of unconventional superconductivity,
Rev. Mod. Phys. {\bf 63}, 239 (1991).   

%\bibitem{tocchio} L.F. Tocchio, A. Parola, C. Gros, and F. Becca, 
%Spin-liquid and magnetic phases in the anisotropic triangular lattice: The case of $\kappa$-(ET)$_2$X, 
%{\prb} {\bf 80}, 064419 (2009).

%\bibitem{Kokalj} J. Kokalj and Ross H. McKenzie, 
%Thermodynamics of a Bad Metal–Mott Insulator Transition in the Presence of Frustration, 
%{\prl} {\bf 110}, 206402 (2013).

%\bibitem{kawakami} T. Yoshioka, A. Koga, and N. Kawakami, 
%Quantum Phase Transitions in the Hubbard Model on a Triangular Lattice,
%{\prl} {\bf 103}, 036401 (2009).

%\bibitem{senechal-spiral} P. Sahebsara and D. S\'en\'echal, 
%Hubbard Model on the Triangular Lattice: Spiral Order and Spin Liquid, 
%{\prl} {\bf 100}, 136402 (2008).

%\bibitem{mila} H.-Y. Yang, A.M. L\"auchli, F. Mila, and K.P. Schmidt, 
%Effective Spin Model for the Spin-Liquid Phase of the Hubbard Model on the Triangular Lattice, 
%{\prl} {\bf 105}, 267204 (2010).

%\bibitem{antipov} A.E. Antipov, A.N. Rubtsov, M.I. Katsnelson, and A.I. Lichtenstein, 
%Electron energy spectrum of the spin-liquid state in a frustrated Hubbard model, 
%{\prb} {\bf 83}, 115126 (2011).


%\bibitem{mishmash} R. V. Mishmash, I. Gonz´alez, R. G. Melko, O. I.
%Motrunich, and M. P. A. Fisher, 
%Continuous Mott transition between a metal and a quantum spin liquid,
%{\prb} {\bf 91}, 235140 (2015).

%\bibitem{misumi}
%K. Misumi, T. Kaneko, and Y. Ohta, 
%Mott transition and magnetism of the triangular-lattice Hubbard model with next-nearest-neighbor hopping,
%Phys. Rev. {\bf B 95}, 075124 (2017).

%\bibitem{szasz} A.~Szasz and J.~Motruk, 
%Phase diagram of the anisotropic triangular lattice Hubbard model, 
%{\prb} {\bf 103}, 235132 (2021). 



%\bibitem{baskaran}G. Baskaran, {\prl} {\bf 100}, 097003 ~(2003).

%\bibitem{ogata}M. Ogata, J. Phys. Soc. Jpn. 72, 1839 ~(2003).





\end{thebibliography}
\end{document}